\documentclass[
pra
preprintnumbers,
 amsmath,amssymb,
 aps,
]{revtex4-2}

\usepackage{graphicx}% Include figure files
\usepackage{dcolumn}% Align table columns on decimal point
\usepackage{bm}% bold math
\usepackage{hyperref}% add hypertext capabilities
\usepackage{amsmath}
\usepackage{algorithm}
\usepackage{algpseudocode}
\usepackage{xr-hyper}

\usepackage{xcolor}
\usepackage{soul}
\setstcolor{black} % Lei
\setulcolor{red} % Lei

\begin{document}

\preprint{revtex4-2}

\title{High-resolution scanning fluorescence imaging through scattering via speckle replica alignment and variance computation}
\author{Lei Zhu$^{1}$}
\author{Tengfei Wu$^{2}$}
\author{Bernhard Rauer$^{1}$}
\author{Hilton B. de Aguiar$^{1}$}
\author{Sylvain Gigan$^{1}$}
\email{Corresponding author: sylvain.gigan@lkb.ens.fr}

\affiliation{$^{1}$Laboratoire Kastler Brossel, ENS--Université PSL, CNRS, Sorbonne Université, Collège de France, 24 Rue Lhomond, F-75005 Paris, France}
\affiliation{$^{2}$State Key Laboratory of Transient Optics and Photonics, Xi’an Institute of Optics and Precision Mechanics, Chinese Academy of Sciences, 710119 Xi’an, China}

\date{\today}% It is always \today, today,

\begin{abstract}
Fluorescence imaging is an essential diagnostic tool in many fields, but diffraction-limited optical imaging at depth is limited by scattering. Here, we present a method based on multiple random illuminations, combined with a computational framework that retrieves high-resolution images by aligning local speckle replicas and computing their pixel-wise variance.  We demonstrate its versatility in two regimes: linear wide-field one-photon (1P) fluorescence imaging and nonlinear two-photon (2P) fluorescence imaging where the object is excited by a scanned speckle field and detected with a single-pixel detector. This approach outperforms standard autocorrelation techniques in terms of resolution and convergence.
\end{abstract}

\maketitle

%\tableofcontents

\section{\label{sec:Introduction}Introduction}

Fluorescent microscopy is an essential tool for biomedical research. However, light transport through an inhomogeneous sample, such as biological tissue, inevitably leads to aberrations and scattering ~\cite{yoon_deep_2020,bertolotti_imaging_2022,gigan_roadmap_2022}. These effects strongly limit the performance of the most advanced fluorescent imaging techniques. To improve image quality in the presence of scattering, conventional fluorescent microscopy enhances performance by exploiting adaptive optics techniques~\cite{hampson_adaptive_2021}, confocal imaging~\cite{wilson_resolution_2011}, or multiphoton imaging~\cite{helmchen_deep_2005}. However, these methods remain fundamentally limited by strong scattering.

Over the past decade, substantial progress has been made in controlling light propagation through scattering media~\cite{yoon_deep_2020, bertolotti_imaging_2022, gigan_roadmap_2022}. Wavefront-shaping approaches actively modulate the incident field—typically with a spatial light modulator (SLM)—to generate a focus behind a scattering layer, using nonlinear fluorescence feedback~\cite{hsieh_imaging_2010,papadopoulos_scattering_2017,berlage_deep_2021,may_fast_2021}, metric-based optimization~\cite{boniface_non-invasive_2019,daniel_light_2019,aizik_fluorescent_2022,aizik_non-invasive_2024}, or transmission-matrix measurements for deterministic focusing~\cite{popoff_measuring_2010,boniface_non-invasive_2020,darco_physics-based_2022,zhao_single-pixel_2024}. Forming a focus alone is insufficient for imaging, but imaging becomes possible when the corrected focus is scanned over a sufficiently large field of view (FOV)—feasible within the optical memory effect~\cite{freund_memory_1988,osnabrugge_generalized_2017}. Alternatively, imaging can be achieved by performing detection-path correction, which compensates scattering-induced aberrations by applying SLM-based phase corrections inferred from guide stars~\cite{katz_looking_2012}, image-quality optimization~\cite{yeminy_guidestar-free_2021}, incoherent iterative phase conjugation~\cite{baek_phase_2023}, or neural-network predictions~\cite{feng_neuws_2023}.

A complementary direction is computational image retrieval without wavefront shaping, exploiting intrinsic speckle correlations~\cite{freund_memory_1988,osnabrugge_generalized_2017}. Speckle-correlation methods reconstruct hidden objects from the autocorrelation of a single speckle image and have been demonstrated in both linear~\cite{bertolotti_non-invasive_2012,katz_non-invasive_2014,hofer_wide_2018,liu_directly_2023} and nonlinear fluorescence regimes~\cite{zhu_two-photon_2025}. However, they require many speckle grains to achieve high-SNR autocorrelations and often rely on iterative phase retrieval—an ill-posed, noise-sensitive procedure with no guaranteed convergence for complex objects~\cite{fienup_phase_1982,wu_single-shot_2016,tian_single-shot_2022}. Other computational approaches based on random illuminations~\cite{wang_non-invasive_2021,zhu_large_2022,soldevila_functional_2023,wu_replica-assisted_2025} have shown promise in linear fluorescence, but are incompatible with nonlinear fluorescence modalities, where the effective point-spread function (PSF) depends on the illumination pattern and therefore cannot be assumed fixed. Recently, matrix-based fluorescence imaging through scattering media~\cite{weinberg_noninvasive_2024} has emerged as a promising alternative; nevertheless, its extension to nonlinear excitation remains largely unexplored. Despite these advances, achieving high-resolution, robust computational image retrieval that operates seamlessly across both linear and nonlinear fluorescence regimes remains a major challenge.

Here, we present a method that allows high-resolution image retrieval from speckle patterns, under unknown random illuminations, valid both linear and nonlinear fluorescence modalities. Our approach leverages local correlations in speckle patterns and their temporal fluctuations, combined with an iterative deconvolution algorithm. 
We experimentally demonstrate the effectiveness of this strategy by imaging fluorescent objects through a scattering medium under unknown random illuminations, both in the linear regime with linear fluorescence and in the nonlinear regime with 2P experiments. A comparison in terms of speed of convergence, image quality and resolution with other techniques showcases the advantage of our method.

\begin{figure*}[ht!]
\includegraphics[width=\textwidth]{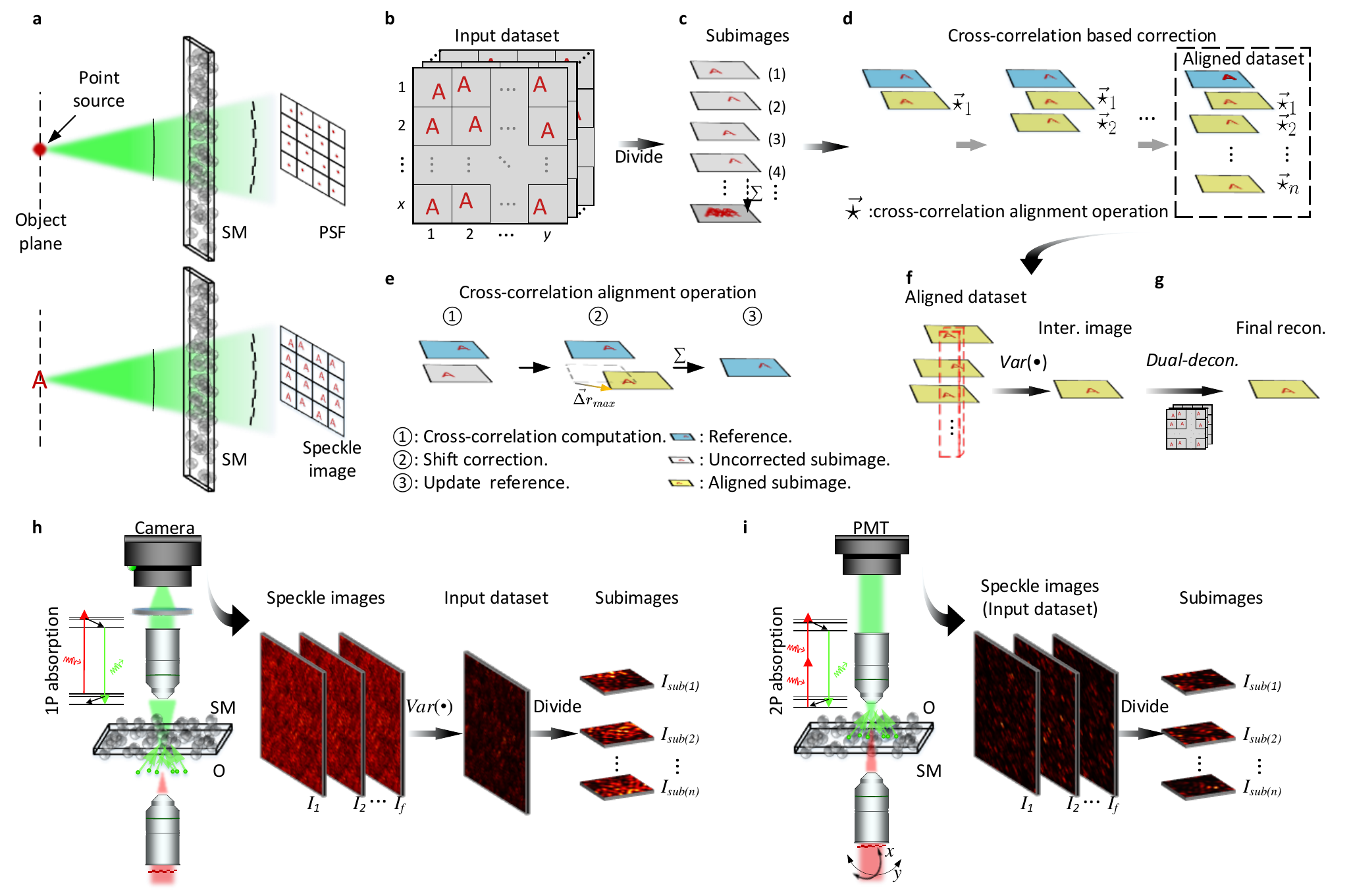}% Here is how to import EPS arts
\caption{\label{fig:1}\textbf{Schematic of the experimental setup and reconstruction principle}. \textbf{a}, When a highly scattering medium obstructs the direct view of an object, light from a point source in the object plane is scattered, creating a diffuse light field and a random speckle on the camera. This speckle pattern can be regarded as the system's PSF. Consequently, the captured speckle image of the object on the camera is the convolution of the object with the PSF.
\textbf{b–f}, The principle of the proposed reconstruction strategy: A series of speckle images is first recorded and then preprocessed according to the specific experimental setup. After preprocessing, the speckle images are transformed into the input dataset (\textbf{b}). This input dataset is then divided into subimages (\textbf{c}), with the size of each subimage determined by the object's approximate dimensions. \textbf{d}, A subimage is selected as a reference, and cross-correlation alignment is performed between the reference and all other subimages. \textbf{e}, During the cross-correlation alignment process: first, the cross-correlation between the reference and an unaligned subimage is computed; second, the unaligned subimage is shifted based on the cross-correlation result; third, the aligned subimage is summed with the reference to update the reference image. \textbf{f}, The intermediate image is achieved by calculating the variance over the aligned data. \textbf{f}, Then, the final reconstruction is produced by performing a dual-deconvolution between this intermediate image and the input dataset.  \textbf{h},\textbf{ 1P fluorescence imaging scenario}: A fluorescent object is illuminated with randomly modulated, unknown laser patterns, and a series of fluorescence speckle images is recorded by the camera as the illumination changes randomly between frames. The pixel-wise variance of the captured speckle images is computed and then used as the input dataset. Subsequently, the input dataset is divided into subimages. \textbf{i}, \textbf{2P fluorescence imaging scenario}: A fluorescent object is excited by a pulsed laser with random, unknown modulation. A speckle image is captured using a photomultiplier tube (PMT) by scanning the speckle pattern across the sample while the rotating diffuser remains fixed during each scan. A series of speckle images is acquired, with the modulation varying between different frames. The captured speckle images act as input data in the 2P fluorescent case. They are then divided into subimages. SM: scattering medium. }
\end{figure*}

\section{\label{sec:Principle}Principle}
\subsection{\label{sec:Imaging_system}Image formation in fluorescence imaging through scattering}

We consider a two-dimensional fluorescent (spatially incoherent) object $O(r)$ imaged through a strongly scattering medium within the angular range of the optical memory effect~\cite{freund_memory_1988,osnabrugge_generalized_2017}. In this regime, image formation can be described by shift-invariant PSFs over a limited field of view.

\textit{General image-formation model.}
We denote by $P_{\mathrm{illum}}(r)$ the speckle pattern describing illumination through the scattering medium and by $P_{\mathrm{det}}(r)$ the corresponding speckle intensity pattern associated with detection. For a fluorescence process of order $p$, the local excitation probability scales as $P_{\mathrm{illum}}^{\,p}(r)$. The detected image can therefore be written in the general form~\cite{bertolotti_imaging_2022,gigan_roadmap_2022,helmchen_deep_2005,zhu_two-photon_2025}
\begin{equation}
I_f(r)=
\begin{cases}
\big[ O(r)\cdot  P_{\mathrm{illum},f}^{\,p}(r) \big] \ast P_{\mathrm{det}}(r), & \text{camera-based detection},\\
O(r) \ast P_{\mathrm{illum},f}^{\,p}(r), & \text{bucket detection},
\end{cases}
\label{EQ:1}
\end{equation}
where $\cdot$ denotes pointwise multiplication, $\ast$ refers to convolution, and the index $f$ labels the $f$-th acquired image corresponding to a distinct speckle realization. 
In the bucket-detection configuration, the speckle illumination pattern is raster-scanned across the object.
This formulation explicitly separates the respective roles of illumination and detection in speckle-based fluorescence imaging.

\textit{1P fluorescence imaging with speckle excitation ($p=1$).}
In widefield 1P fluorescence microscopy, illumination is uniform or weakly structured, and fluorescence emission propagates through the scattering medium before being recorded by a camera. The effective PSF is therefore imposed by the detection speckle. For a single exposure, the recorded image is given by
\begin{equation}
I_{\mathrm{1P},f}(r) = \big[ O(r)\cdot P_{\mathrm{illum},f}(r) \big] * P_{\mathrm{det}}(r),
\label{EQ:2}
\end{equation}
corresponding to a single-shot camera image within the memory-effect range.

\textit{2P fluorescence imaging with speckle excitation ($p=2$).}
In 2P fluorescence imaging, the sample is excited by a speckle illumination pattern rather than a diffraction-limited focus~\cite{zhu_two-photon_2025}. Owing to the quadratic dependence of two-photon excitation, the effective excitation profile is given by $P_{\mathrm{illum}}^{\,2}(r)$. The resulting fluorescence is collected by a non--spatially resolved bucket detector, typically a photomultiplier tube (PMT).

Because detection is spatially integrated, the effective PSF is determined solely by the illumination pattern. For a single speckle realization, the reconstructed image obtained by raster-scanning the speckle pattern is therefore
\begin{equation}
I_{\mathrm{2P},f}(r) = O(r) \ast P_{\mathrm{illum},f}^{\,2}(r).
\label{EQ:3}
\end{equation}
This scheme is conceptually analogous to conventional point-scanning 2P microscopy~\cite{helmchen_deep_2005}, with the key distinction that excitation is mediated by a random speckle pattern. The framework naturally extends to higher-order nonlinear excitation, such as three-photon (3P) fluorescence imaging.

Despite the differences in illumination and detection pathways in the 1P and 2P regimes, both image-formation models reduce to an effective convolution with a speckle-dependent kernel. The reconstruction relies exclusively on disorder-induced correlations encoded in the measured images, arising from multiple embedded replicas of the object within each speckle image, which are preserved across these models.
\subsection*{\label{sec:Reconstruction_algorithm}Reconstruction algorithm}

Since the hidden object produces many randomly distributed replicas on the camera, the captured image is of low contrast and bears little resemblance to the object itself. Although, in principle, recovering any single replica would be sufficient to retrieve the object’s spatial information, isolating one instance without interference from the others is extremely challenging. Instead, we capitalize here on the presence of multiple embedded replicas across the field of view. By exploiting the correlations shared among these subregions—each containing only a few shifted instance of the same object—we reconstruct the object without explicitly identifying any individual replica. Our reconstruction framework comprises four stages: (a) subimage division, (b)cross-correlation–based correction, (c) pixel-wise variance computation, and (d) dual deconvolution (Fig.~\ref{fig:1}b–g). 

More in detail, starting from the preprocessed input dataset (Fig.~\ref{fig:1}b), we (a) partition the measurement into a set of partially overlapping subimages (Fig.~\ref{fig:1}c). The subimage size is selected based on an estimate of the object’s spatial extent derived from autocorrelation analysis. Each subimage is then (b) refined through an iterative cross-correlation–based correction procedure using a reference pattern (Fig.~\ref{fig:1}d–e). In each iteration, we compute the normalized cross-correlation~\cite{lewis1995fncc} between the reference $R_n$ and a subimage $S_n$:

\begin{equation}
\begin{split}
 &C_{n}(\vec{\Delta r}) = \\ &\frac{
\sum_{\mathbf{r}} \left( R_{n}(\mathbf{r}) - \bar{R}_{n} \right) \left( S_{n}(\mathbf{r} + \vec{\Delta r}) - \bar{S}_{n, \vec{\Delta r}} \right)
}{
\sqrt{
\sum_{\mathbf{r}} \left( R_{n}(\mathbf{r}) - \bar{R}_{n} \right)^2
\cdot
\sum_{\mathbf{r}} \left( S_{n}(\mathbf{r} + \vec{\Delta r}) - \bar{S}_{n, \vec{\Delta r}} \right)^2
}
}.
\end{split}
\label{EQ.4}
\end{equation}

The displacement that maximizes the correlation,

\begin{equation}
\vec{\Delta r}_{\text{max}} = \arg \max_{\vec{\Delta r}} \, C_{n}(\vec{\Delta r}), 
\label{EQ.5}
\end{equation}
is applied to spatially register the subimage:
\[
S_{n}^{\text{aligned}} = S_{n}(\mathbf{r} + \vec{\Delta r}_{\text{max}}).
\]

The reference is then updated through a correlation-weighted addition:
\begin{equation}
R_{n+1} = R_{n} + \max_{\vec{\Delta r}} C_n  S_{n}^{\text{aligned}}.
\label{EQ.6}
\end{equation}
By repeating this for every subimage, we build a highly accurate, perfectly overlaid set of images $S_{n}^{\text{aligned}}$.

Once all subimages are well registered, (c) we compute the pixel-wise variance across the aligned ensemble (Fig.~\ref{fig:1}f):
\begin{equation}
Inter(r) = Var_{n}\left( S_{n}^{\text{aligned}} \right).
\label{EQ.7}
\end{equation}
This operation yields an intermediate representation that effectively enhances spatial detail by exploiting the temporal fluctuations in the aligned data while suppressing background noise.
(d) A final dual-deconvolution between this intermediate image and the input dataset produces the final reconstruction $Recon(r)$ (Fig.~\ref{fig:1}g). Full algorithmic details are provided in the Methods \ref{Workflow of the Proposed Method}.

The preprocessing differs between 1P and 2P fluorescence imaging due to their distinct excitation physics. In 1P imaging, fluorescence scales linearly with illumination intensity, resulting in characteristic speckle statistics; thus we first compute the pixel-wise variance of the raw measurements and use this as the input dataset (Fig.~\ref{fig:1}h). In contrast, 2P fluorescence exhibits a quadratic intensity dependence and higher speckle contrast, allowing the raw frames themselves to serve directly as the input dataset for subimage extraction (Fig.~\ref{fig:1}i).

\begin{figure*}
\includegraphics[width=\textwidth]{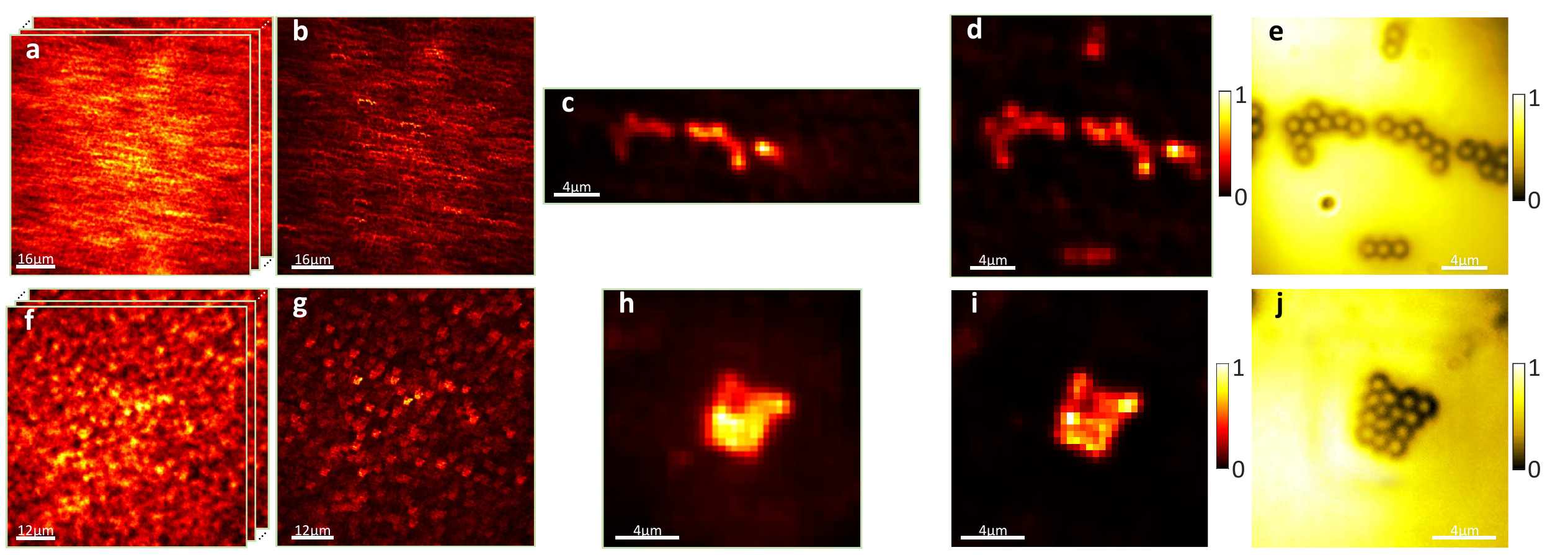}% Here is how to import EPS arts
\caption{\label{fig:2}\textbf{1P fluorescence imaging through scattering media}. \textbf{a-f}, Fluorescence speckle images measured at the camera. \textbf{b-g}, Input data, which is preprocessed data from (\textbf{a-f}). \textbf{c-h}, The intermediate image. \textbf{d-i}, The final reconstruction. \textbf{e-j}, Ground truth image taken from the side without the scattering media.}
\end{figure*}

\section{\label{sec:Results}Results}
To demonstrate the feasibility of our methods, we have conducted experiments under both 1P and 2P fluorescent imaging conditions. In each case, fluorescent beads are placed behind a highly scattering layer and excited using random illumination. Subsequently, the scattered fluorescent signal is captured by the detector. Further experimental details are provided below.

\subsection{1P fluorescent imaging validation}

To validate our method in 1P fluorescence imaging, fluorescent beads were used as the sample and a continuous-wave (CW) laser served as the illumination source. The wavefront was modulated using a SLM positioned conjugate to the back focal plane of the illumination objective. The beads were drop-cast onto the bottom surface of a glass coverslip facing the illumination objective, while the top surface was sand-blasted to introduce a strongly scattering layer.

The modulated illumination excites 1P fluorescence, which passes through the scattering medium and produces a scattered speckle image. The speckle images were collected using a detection objective and tube lens and recorded by an sCMOS camera (see Methods for experimental setup~\ref{1P experimental setup}). A series of speckle images was captured under random SLM patterns (Fig.~\ref{fig:2}a), and the pixel-wise variance across the dataset was computed to form a single frame for reconstruction (Fig.~\ref{fig:2}b).

The procedure above was implemented: the input dataset was divided into multiple subimages, with the size of each subimage estimated via autocorrelation of the data \cite{bertolotti_non-invasive_2012}. A cross-correlation-based alignment procedure (Fig.~\ref{fig:1}d) was applied to the subimages to produce aligned subimages. Computing the pixel-wise variance over the aligned subimages yielded the intermediate image of the hidden object (Fig.~\ref{fig:2}c). Dual-deconvolution between the input dataset and the intermediate image produced the final reconstruction (Fig.~\ref{fig:2}d), which closely matches the ground-truth image obtained without scattering media (Fig.~\ref{fig:2}e). These results demonstrate the effectiveness of our method for 1P fluorescence imaging through scattering media.

\begin{figure*}[t!]
\includegraphics[width=\textwidth]{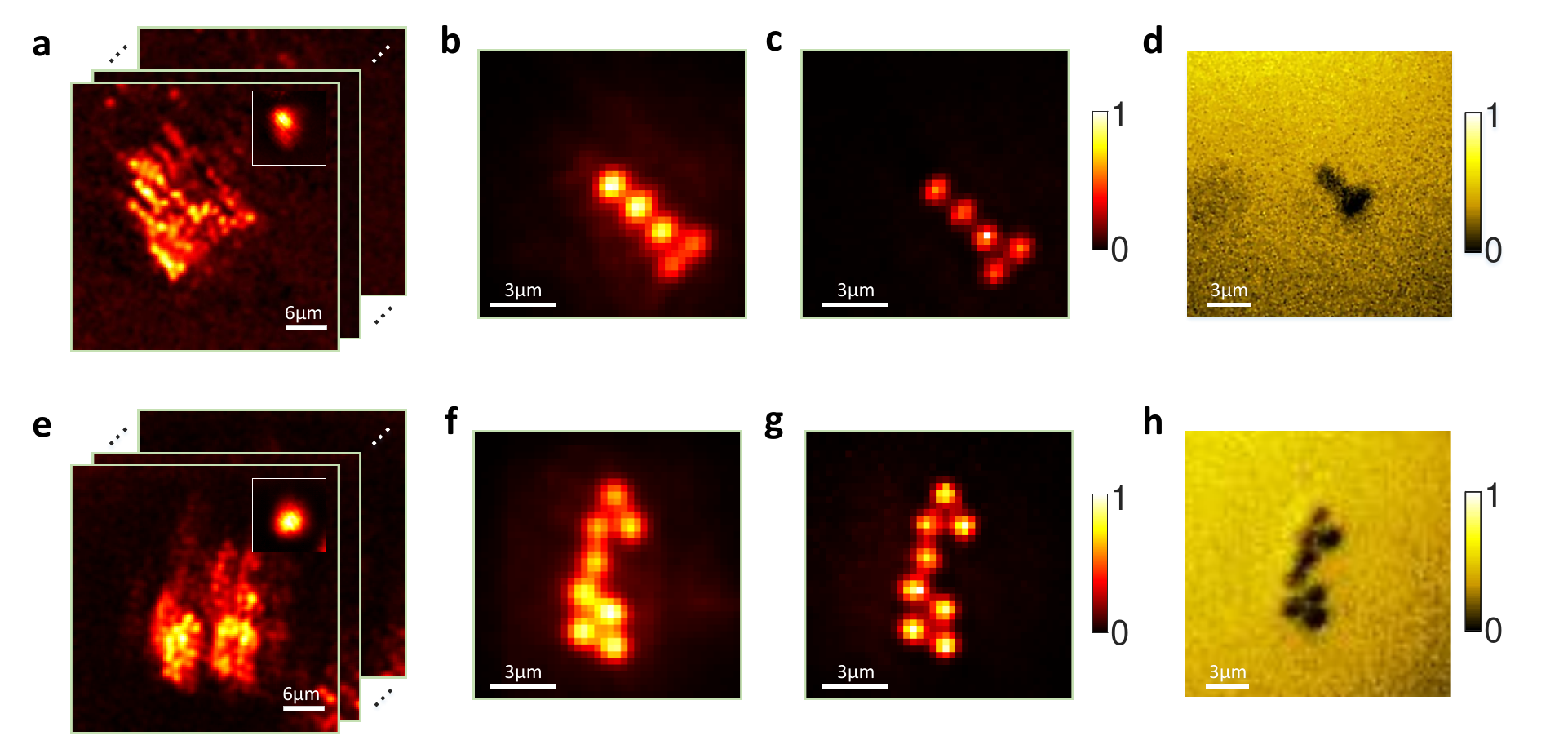}% Here is how to import EPS arts
\caption{\label{fig:3} \textbf{2P fluorescence imaging through scattering media}. \textbf{a-e}, Fluorescence speckle images captured using PMT. (Inset) Incoherent sum of speckle images. \textbf{b-f}, The intermediate image. \textbf{c-g}, The final reconstruction. \textbf{d-h}, Ground truth images taken from the side without the scattering media.} 
\end{figure*}

\subsection{2P fluorescent imaging validation}

To validate our method in 2P fluorescence imaging, the same fluorescent beads were used as the sample and a femtosecond pulsed laser (see methods section \ref{2P experimental setup})  was employed as the illumination source. Random illumination patterns were generated with a rotating diffuser positioned conjugate to the back focal plane of the illumination objective. The beads were drop-cast onto the top surface of a glass coverslip, while the bottom surface facing the illumination objective was sand-blasted to introduce a scattering layer.

The speckled laser illumination excites 2P fluorescence, which is collected by an objective and detected using a PMT. Individual 2P speckle images were acquired by scanning the speckle pattern across the sample using galvo mirrors, while keeping the diffuser stationary, analogous to standard point-scanning 2P microscopy. Different speckle realizations were obtained by rotating the diffuser between acquisitions (Fig.~\ref{fig:3}a)(see Methods for experimental setup~\ref{2P experimental setup}). Unlike 1P imaging, the raw 2P speckle images were used directly as the input dataset without preprocessing.

The reconstruction pipeline was then applied in the same manner as for 1P: the dataset was divided into subimages, with subimage size determined via autocorrelation of the speckle data. Cross-correlation-based alignment produced aligned subimages, and computing the pixel-wise variance over these aligned subimages generated the intermediate image of the hidden object (Fig.~\ref{fig:3}b). Dual-deconvolution between the input dataset and the intermediate image produced the final reconstruction (Fig.~\ref{fig:3}c), which closely matches the ground-truth image obtained without scattering media (Fig.~\ref{fig:3}d). These results validate the effectiveness of our method for 2P fluorescence imaging through scattering media.

\begin{figure*}[t!]
\includegraphics[width=\textwidth]{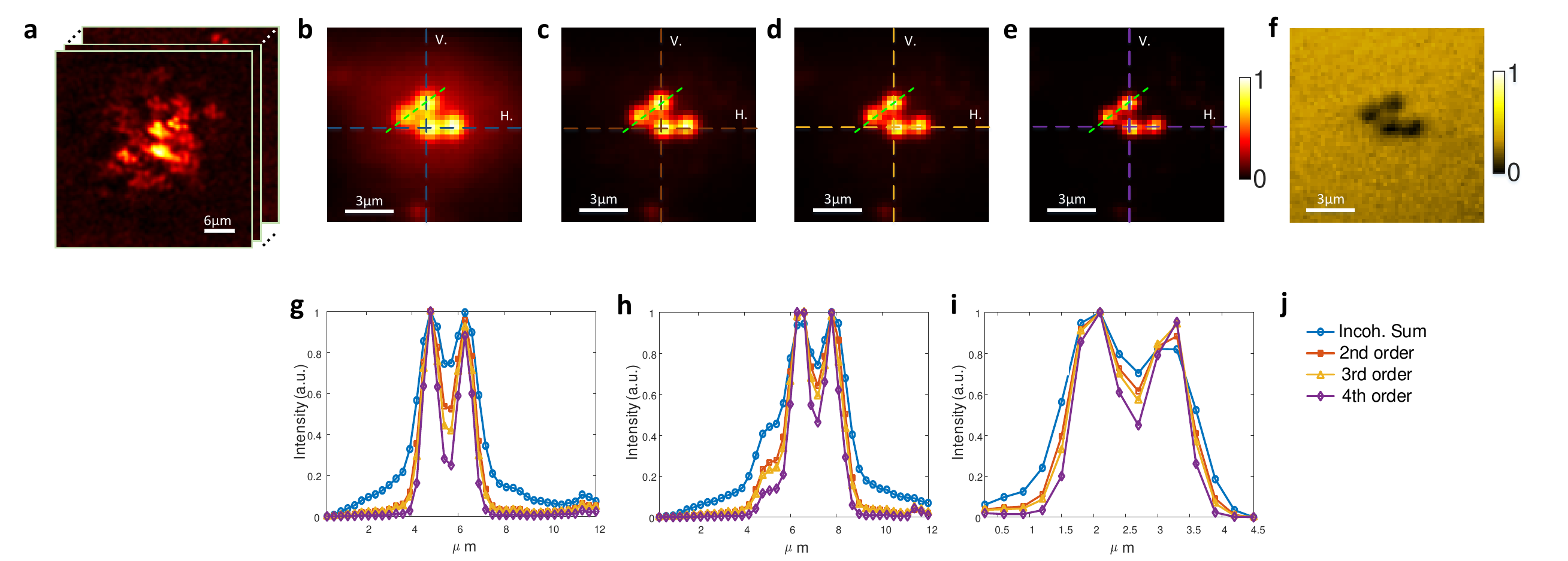}
\caption{\label{fig:4} \textbf{Comparison of resolution between incoherent summation and higher-order SOFI over aligned subimages.} 
\textbf{a}, 2P fluorescence speckle images captured using a PMT. 
\textbf{b–e}, Incoherent sum, and second-, third-, and fourth-order SOFI images, respectively. 
\textbf{f}, Ground-truth images taken from the side without scattering media. 
\textbf{g–h}, Intensity profiles extracted along the dashed lines in \textbf{b–e}. 
\textbf{g}, Vertical profiles; \textbf{h}, Horizontal profiles. \textbf{i} Intensity profiles extracted along the green dashed lines in \textbf{b–e}. \textbf{j}, Legends for \textbf{g–i}.}
\end{figure*}

A key feature of our method is its high-resolution capability, enabled by the pixel-wise variance computation. This step is conceptually analogous to super-resolution optical fluctuation imaging (SOFI)~\cite{dertinger_fast_2009, basak_super-resolution_2025}, in which spatial intensity fluctuations reveal information beyond conventional incoherent summation (see Methods section for further details~\ref{Rationale for pixel-wise variance computation}). To illustrate this effect, we performed an additional 2P imaging experiment and processed the data using our reconstruction pipeline, omitting only the final dual-deconvolution step. Incoherent summation of aligned subimages produces a resolution-limited image (Fig.~\ref{fig:4}b), whereas second- to fourth-order SOFI images resolve finer details (Fig.~\ref{fig:4}c–e), consistent with the ground-truth image (Fig.~\ref{fig:4}f). Line profiles (Fig.~\ref{fig:4}g–i) confirm that high-order correlation enhances resolution beyond incoherent summation, validating the SOFI-like mechanism.

\section{\label{sec:Discussion}Discussion and conclusion}

We have presented a method that enables high-resolution reconstruction of hidden objects from speckle patterns under unknown random illuminations, applicable to both linear (1P) and nonlinear (2P) fluorescence modalities. Our strategy avoids the challenges associated with phase retrieval algorithms, and intensive wavefront optimization methods, while providing more reliable reconstructions for complex objects compared with traditional autocorrelation-based approaches~\cite{bertolotti_non-invasive_2012, katz_non-invasive_2014, hofer_wide_2018, zhu_two-photon_2025}(see Methods section for numerical simulation~\ref{Comparison With autocorrelation-based reconstruction}).

As a downside, we note that our method requires a dual-deconvolution step. In principle, with a sufficiently large number of subimages, pixel-wise variance alone can enable reconstruction, as shown in Fig.~\ref{fig:4}. In practice, however, limited data prevent pixel-wise variance computation from attaining the optimal resolution of the method. The dual-deconvolution step aggregates information from multiple replicas, thereby improving both reconstruction quality and achievable resolution under realistic experimental conditions.

An additional advantage of our approach is that engineered or naturally induced disorder generates multiple embedded replicas of the object, which serve as intrinsic correlation references. Crucially, these correlations are preserved across different scattering imaging modalities due to the convolutional nature of the image formation, rendering the method inherently modality-agnostic and applicable to a wide range of scattering imaging scenarios.
The fidelity of the recovered image depends on the statistical properties of the effective PSF, particularly its contrast and background suppression, which can be enhanced through preprocessing when required. A key practical limitation is that the accessible FOV is governed by the angular range of the optical memory effect. In moderately scattering media, where this range is reduced, the field of view can be extended by sampling multiple isoplanatic regions sequentially, with each measurement confined to a single memory-effect patch~\cite{sunray_beyond_2024}.

Overall, our work establishes a versatile computational imaging framework capable of recovering hidden objects in both linear and nonlinear regimes under highly scattering conditions. With sufficient data, higher-order correlation computations could further enhance resolution, and the approach is readily extendable to other imaging modalities, including 3P fluorescence. We anticipate that this method will open new avenues for computational imaging and sensing applications in complex scattering environments using spatially incoherent light.

%\appendix
\section{\label{sec:Methods}Methods}
\subsection{\label{Experimental setup}Experimental setup}
\subsubsection{\label{1P experimental setup}1P experimental setup}
The experimental schematic is presented in Fig.~\ref{fig.5}a. A $532~ \ nm$ CW laser is used to excite the fluorescent signal from the sample consisting of fluorescent beads (FluoSpheres\textsuperscript{TM} carboxylate, $1.0\,\mu m$, 540/560), dropcasted on the bottom surface of a 1-mm-thick microscope slide. The opposite side is sandblasted to induce a single scattering layer with 220 grit sand. A SLM (HSP1920532-HSP8, Meadowlark Optics) is used, and various random phase patterns are sequentially sent to the SLM to generate multiple speckle realizations at the object plane. The modulated excitation is focused by an objective (CFI Plan Fluor 60$\times$/0.85, Nikon). A beamsplitter with a 2/98 transmission/reflection ratio is used to monitor the excitation speckle illumination and to inspect the ground truth via free-space brightfield imaging using a CMOS camera (acA4024-29um, Basler) and a tube lens (TL, $f = 150\, mm$). The emitted fluorescent speckle patterns are collected by an objective (Plan N 10$\times$/0.25, Olympus), and recorded on an sCMOS camera (Flash v4.0, Hamamatsu) with a TL ( $f = 300\,mm$). A set of filters—a bandpass filter (FBH560-10, Thorlabs), a longpass filter (FELH0550, Thorlabs), and a notch filter (NF533-17, Thorlabs)—are used to sufficiently isolate the fluorescence signal.

\subsubsection{\label{2P experimental setup}2P experimental setup}

\begin{figure*}[t!]
\includegraphics[width=0.8\textwidth]{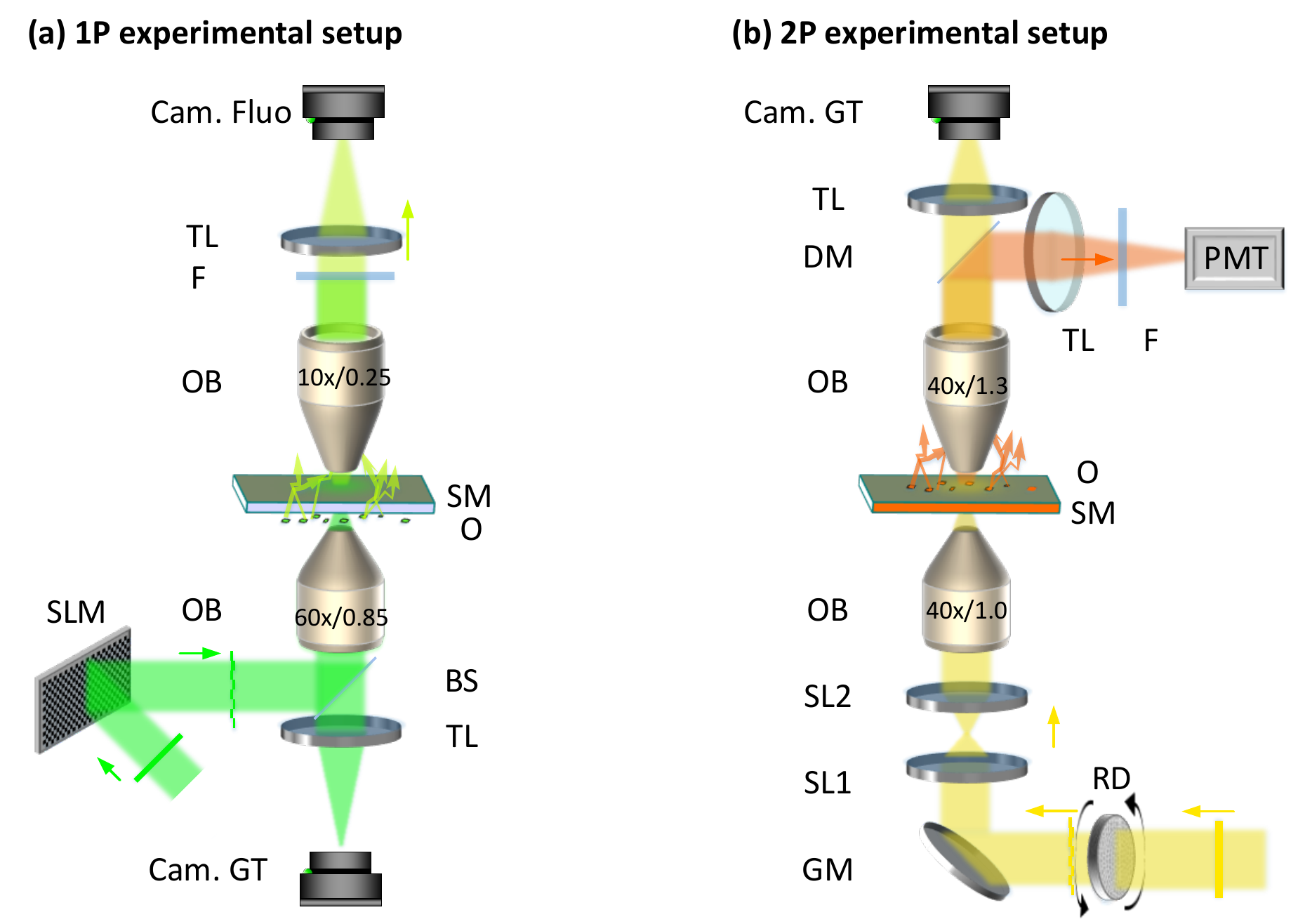}
\caption{\textbf{Experimental setups.}
(a) \textbf{ 1P Setup:} A sequence of random phase patterns is displayed on a spatial light modulator (SLM), generating uncorrelated speckle illumination patterns at the sample plane (O). This is achieved through a beam splitter (BS) and a high-numerical-aperture objective (OB) with a NA of 0.85, which focuses the light onto fluorescent beads. The emitted fluorescence is transmitted through a scattering medium (SM) and collected by a low-NA OB (NA = 0.25). The signal is then imaged onto a fluorescence camera (Cam. fluo) via a tube lens (TL). The camera ground truth (Cam. GT) is used to capture the ground truth image from the side without SM.\\
(b) \textbf{ 2P Setup:} A femtosecond laser beam is modulated by a rotating diffuser (RD), introducing random phase variations. The modulated beam is directed through a galvanometric mirror (GM) unit for raster scanning and focused onto the back focal plane of an excitation OB (NA = 1.0), exciting the object located behind a SM. The emitted fluorescence is collected by a second OB (NA = 1.3), filtered, and detected by a photomultiplier tube (PMT) after passing through optical components including a dichroic mirror (DM), tube lens (TL), and filter (F). The Cam. GT is operated to capture the ground truth image from the side without SM.\\ The beam shown in green in \textbf{(a)} (in yellow in \textbf{(b)}) corresponds to the illumination light, whereas the beam shown in lime yellow in \textbf{(a)} (in vivid orange in \textbf{(b)}) corresponds to the emission light.
\textbf{Abbreviations:} SLM – spatial light modulator; BS – beam splitter; OB – objective; SM – scattering medium; CAM – camera; PMT – photomultiplier tube; GM – galvo mirror; SL – scanning lens; DM – dichroic mirror; TL – tube lens; F – filter; O – object; Cam. GT - camera ground truth.
}
\label{fig.5}
\end{figure*}

The experimental schematic is presented in Fig.~\ref{fig.5}b. A pulsed Ti:sapphire laser (Chameleon Ultra II, Coherent) pumps an optical parametric oscillator (OPO-X, Coherent Mira), generating a two-photon (2P) excitation source centered at $1050\,nm$ with a pulse duration of $140\,fs$ and a repetition rate of $80\,MHz$.
The raster scanning system consists of a pair of galvanometric mirrors and a 4$f$ optical relay, conjugated to the back focal plane of a water-immersion objective (W Plan-Apochromat 40$\times$ /1.0 DIC M27, Zeiss). Fluorescence emission is collected by a second objective (EC Plan-NEOFLUAR 40$\times$/1.3, Zeiss) positioned opposite the first objective.
The collected light is directed through a dichroic mirror (Di03-R785-t1-25×36, Semrock), passed through an emission filter, and detected by a PMT (H7422P-40, Hamamatsu). During sample preparation, the second objective is also used to image the sample onto a CMOS camera (acA1300-30$\mu$m, Basler) for positioning and alignment.
The scattering medium is created by sandblasting the bottom surface of a No.~1.5H coverslip to produce a rough surface with 220 grit sand. Fluorescent beads (FluoSpheres\textsuperscript{TM} carboxylate, $1.0\,\mu m$, 580/605) are dropcasted onto the top surface of the coverslip, approximately $340~\mu m$ above the scattering layer. 
To achieve a near-index-matched environment, the sample is encapsulated in NOA85 optical adhesive. A second No.~1.5H coverslip is placed over the sample, and the adhesive is cured under ultraviolet light.

\subsection{\label{Experimental parameters}Experimental parameters}
The experimental parameters for the results displayed in Figs.~\ref{fig:2}-\ref{fig:4}, including image pixel count, number of images, overlap ratio between subimages, and the scatterer’s location, are outlined as follows: In Fig.~\ref{fig:2}a(b), images are $580 \times 580$ ($372 \times 384$) pixels with a $100\,ms$ camera exposure, the number of images is 150, the overlap ratio between the subimages is $40\%$, and the scattering layer is inserted between the detection objective and its focal plane. In Figs.~\ref{fig:3}a,e and Fig.~\ref{fig:4}a, images are $128 \times 128$ pixels with a $100\,\mu s$ pixel dwell time, the number of images is 80 in Figs.~\ref{fig:3}a,e and Fig.~\ref{fig:4}a, the overlap ratio between the subimages is $30\%$, and the scattering layer is located between the illumination objective and its focal plane. The value of $\alpha$ in the dual-deconvolution step is 18 in Fig.~\ref{fig:2}d,i and 2 in Fig.~\ref{fig:3}c,g, respectively.

\subsection{\label{Rationale for pixel-wise variance computation}Rationale for pixel-wise variance computation}

The $n$-th aligned subimage, $S_{n}^{\mathrm{aligned}}(r)$, can be expressed as a superposition of multiple object replicas:
\begin{equation}
S_{n}^{\mathrm{aligned}}(r)
= \sum_{i=1}^{M} a_{i}\,\big[ O(r-\delta r_{i}) \ast kernel_{i} \big], \qquad a_i \le 1,
\label{EQ:8}
\end{equation}
where $M$ denotes the total number of object replicas contained in $S_{n}^{\mathrm{aligned}}$, $a_i$ is the relative illumination coefficient of the $i$-th replica, $\delta r_i$ represents its relative spatial shift, $ \ast$ corresponds to the convolution operator, and $kernel_i$ indicates the corresponding convolution kernel, which is given by an individual speckle grain.

We assume that, for each aligned subimage, a single replica dominates the local intensity. Without loss of generality, we re-index this dominant replica such that $\delta r_{\mathrm{dominant}} = 0$. Equation~\eqref{EQ:8} can then be rewritten as
\begin{align}
S_{n}^{\mathrm{aligned}}(r)
&= a_{\mathrm{dominant}}\,\big[ O(r) \ast kernel_{\mathrm{dominant}} \big] \nonumber \\
&\quad + \sum_{i=1}^{M-1} a_i\,\big[ O(r-\delta r_i) \ast kernel_i \big],
\qquad a_i < a_{\mathrm{dominant}} .
\label{EQ:9}
\end{align}

In Eq.~\eqref{EQ:9}, the first term corresponds to the dominant replica, while the second term represents multiple dimmer, spatially shifted replicas convolved with distinct speckle kernels. Owing to the random nature of scattering, these dim replicas partially overlap with the dominant replica. Their combined contribution can therefore be regarded as spatially varying local fluctuations superimposed on the dominant signal.

Under this approximation, Eq.~\eqref{EQ:9} can be recast into the effective form
\begin{equation}
S_{n}^{\mathrm{aligned}}(r)
= Fluc._n(r)\,\cdot\,\big[ O(r) \ast kernel_{\mathrm{dominant}} \big]
+ Noise_n(r),
\label{EQ:10}
\end{equation}
where $Fluc._n(r)$ denotes an equivalent spatially varying multiplicative fluctuation field acting on the dominant replica, and $Noise_n(r)$ accounts for all remaining contributions that are not correlated with the dominant replica. Here, $\cdot$ denotes pointwise multiplication.

Due to the ergodic-like properties of speckle in the diffusive regime~\cite{freund_looking_1990}, the overlap between the dominant and weaker replicas is statistically uncorrelated and randomly distributed across different subimages. As a result, the fluctuation patterns at the location of the dominant replica in each aligned subimage are statistically independent from those in other aligned subimages.
Overall, the second-order correlation, equivalent to the pixel-wise variance computation, can be leveraged to reconstruct the object while effectively suppressing the noise component across a stack of aligned subimages. 
The second-order correlation is given by:

\begin{equation}
G^{(2)}(r) = \left\langle \left(S_{n}^{\text{aligned}}(r)\right)^2 \right\rangle_n - \left\langle S_{n}^{\text{aligned}}(r) \right\rangle_n^2,
\label{EQ.11}
\end{equation}
where $\langle \cdot \rangle_n$ denotes the averaging over all subimages indexed by $n$. This operation selectively amplifies the object signal convolved with the dominant kernels, while attenuating uncorrelated noise and randomly distributed contributions from weaker replicas.

The corresponding pixel-wise variance across $N$ aligned subimages can be expressed as:

\begin{align}
Var_n\left(S_{n}^{\text{aligned}}(r)\right)
&= \left| O(r) \ast \left\langle{kernel}_{\text{dominant}} \right\rangle_{n} \right|^2 \cdot Var_{n}\left( \text{Fluc.}_{n}(r) \right) \nonumber \\
&\quad + Var_{n}\left(Noise_{n}(r) \right).
\label{EQ.12}
\end{align}
 
In the regime where the fluctuation-induced signal dominates over the noise—such as in high speckle contrast conditions or with sufficient ensemble averaging—the noise term becomes negligible:

\begin{equation}
Var_{n}\left(S^{\text{aligned}}(r)\right) \approx \left| O(r) \ast  \left\langle{kernel}_{\text{dominant}} \right\rangle_{n} \right|^2 \cdot Var_{n}\left( \text{Fluc.}_{n}(r) \right).
\label{EQ.13}
\end{equation}

If $Var_{n}(\text{Fluc.}_{n}(r))$ is assumed to be spatially uniform or normalized (e.g., set to unity), the variance expression simplifies to:

\begin{equation}
Var_{n}\left(S_{n}^{\text{aligned}}(r)\right) \propto \left| O(r) \ast  \left\langle{kernel}_{\text{dominant}} \right\rangle_{n} \right|^2,
\label{EQ.14}
\end{equation}
providing a second-order statistical estimate of the object, convolved with the squared average kernel of the dominant replica.

\subsection{\label{Dual-deconvolution}Dual-deconvolution}

The dual-deconvolution procedure comprises two successive deconvolution stages. In the first stage, deconvolution is performed frame-wise between the intermediate image $Inter(r)$ and each input frame $Input_{m}(r)$, expressed as

\begin{equation}
Speckle_{m}(r) = Deconvol(Input_{m}(r),Inter(r)).
\label{EQ.15}
\end{equation}
Here, $m$ indicates the frame index of $Inter(r)$.

Next, the lowest $\alpha\%$ of the pixel intensities in the resulting speckle pattern $Speckle_{m}$ are set to zero, yielding $Speckle_{m}^{threshold}$. The value of $\alpha$ ranges from 2 to 20. In the second stage, the deconvolution is performed frame-wise between this thresholded speckle $Speckle_{m}^{threshold}$ and the input dataset $Input_{m}(r)$, and the final reconstruction is obtained by averaging over all frames:

\begin{equation}
Recon(r) = \left\langle Deconvol(Input_{m}(r),Speckle_{m}^{threshold}(r))\right\rangle_{m},
\label{EQ.16}
\end{equation}
where $\langle \cdot \rangle_m$ denotes averaging over the frame index $m$.
In our work, a standard Lucy–Richardson deconvolution is employed \cite{biggs_acceleration_1997}.

\subsection{\label{Workflow of the Proposed Method}Workflow of the Proposed Method}

This section outlines the post-processing pipeline used to recover the image of the hidden object from a series of recorded speckle images. 
The reconstruction process consists of five main steps, as explained follows:

\textbf{Step 1: Imaging Mode Selection.}  
Depending on the imaging modality—either 1P or 2P —the preprocessing of the raw speckle images $ \{I_f\}_{f=1}^F $ differs. For the 1P case, a pixel-wise variance is computed across the full stack of speckle images to enhance object features and suppress background noise. In contrast, for the 2P case, the raw speckle images are used directly as input without initial variance computation.
\textbf{Step 2: Subimage Segmentation.}  
The input image, denoted as $ \text{Input}(r) $, is segmented into $ N $ overlapping or non-overlapping subimages $ \{S_n\}_{n=1}^N $, based on an estimated object size or feature scale. 
%\ul{This facilitates localized alignment and improves reconstruction quality by reducing spatial variance across larger fields of view.}
\textbf{Step 3: Reference Selection.}  
One of the subimages, typically the one with the highest contrast, is selected as the initial reference $ R_1 $. This reference guides the alignment of other subimages in the following steps.
\textbf{Step 4: Cross-Correlation-Based Alignment.}  
An iterative alignment process is performed over a predefined number of trials. For each subimage $ S_k $, the normalized cross-correlation $ C_k $ with the current reference $ R_k $ is computed. If the maximum correlation $ \max_{\vec{\Delta r}} C_k $ exceeds a threshold (e.g., 0.2), a displacement vector $ \vec{\Delta \mathbf{r}}_{\text{max}} $  is estimated from the peak of the correlation map. The subimage is then shifted accordingly to produce an aligned version $ S_k^{\text{aligned}} $. The reference image is updated by accumulating the weighted contribution of the aligned subimage. If the correlation is below the threshold, the subimage is skipped, and the reference remains unchanged.
\textbf{Step 5: Variance computation.}  
After all subimages are aligned, the pixel-wise variance is computed across the aligned stack $ \{S_k^{\text{aligned}}\}_{k=1}^K $. This variance operation enhances consistent structural features while suppressing uncorrelated noise, yielding the intermediate image $ \text{Inter}(r) $.
\textbf{Step 6: Dual-deconvolution.}  
The dual-deconvolution is performed between the intermediate image $ \text{Inter}(r) $ and the input dataset to generate the final reconstruction $ \text{Recon}(r) $.
The complete workflow is summarized as follows:

\begin{algorithm}[H]
\label{alg:reconstruction}
\caption{Step-by-step reconstruction procedure}
\begin{algorithmic}[1]
\Require Stack of $F$ speckle images: $\{I_f\}_{f=1}^{F}$
\Ensure Reconstructed image $Recon(r)$

\State \textbf{Step 1: Choose imaging mode (1P or 2P)}
\If{1P case}
    \State Compute pixel-wise variance directly:
    \[
    Input(r) = Var_f\left(I_f(r)\right)
    \]
\ElsIf{2P case}
    \State Use raw speckle images as input: $Input(r) \gets I_f(r)$
\EndIf

\State \textbf{Step 2: Divide image into subimages}
\State Segment $Input(r)$ into $N$ subimages based on estimated object size:
\[
\{S_n\}_{n=1}^{N}
\]

\State \textbf{Step 3: Reference selection}
\State Select one subimage as the initial reference $R_{1}$ (e.g., with highest contrast)
\State \textbf{Step 4: Cross-correlation based correction}
\For{$trial = 1$ to $NUM$}
\State t = 0
\For{$k = 1$ to $N$}
    \State t = t+1
    \State Compute normalized cross-correlation $C_k$ between $S_k$ and $R_k$
    \If{ $\max_{\vec{\Delta r}} \, C_{n} \geq 0.2$}
    \State Estimate shift $\vec{\Delta \mathbf{r}}_{\text{max}}$ from correlation peak
    \State Align subimage:
    \[
    {S}^{aligned}_k(r) = S_k(\mathbf{r} + \vec{\Delta r}_{\text{max}})
    \]
    \State Update reference image:
    \[
    R_{t+1} \gets R_t + \max_{\vec{\Delta \mathbf{r}}} \, C_k \cdot S_k(\mathbf{r} + \vec{\Delta \mathbf{r}}_{\text{max}})
    \]
    \ElsIf{ $\max_{\vec{\Delta r}} \, C_{n} < 0.2$}
    \State Skip alignment of subimage:
    \State Update reference image:
    \[
    R_{t+1} \gets R_t
    \]
    \EndIf
\EndFor
\EndFor

\State \textbf{Step 5: Variance computation}
\State Stack all aligned subimages ${S}^{aligned}_k(r)_{k=1}^{K}$
\State Compute pixel-wise variance over aligned subimages:
\[
Inter(r) = Var_k\left({S}^{aligned}_k(r)\right)
\]
\State \textbf{Step 6: Dual-deconvolution}
\State Dual deconvolution between $Input(r)$ and $Inter(r)$ :
\[
Recon(r) = \text{dual-deconvolutin}\left( Input(r),Inter(r) \right)
\]
\State \Return $Recon(r)$
\State \textbf{Note:} Parameter $NUM$ is set to $4$ in all experimental reconstructions.
\end{algorithmic}
\end{algorithm}

\subsection{\label{Comparison With autocorrelation-based reconstruction}Comparison with autocorrelation-based reconstruction}

\begin{figure*}[t!]
\includegraphics[width=\textwidth]{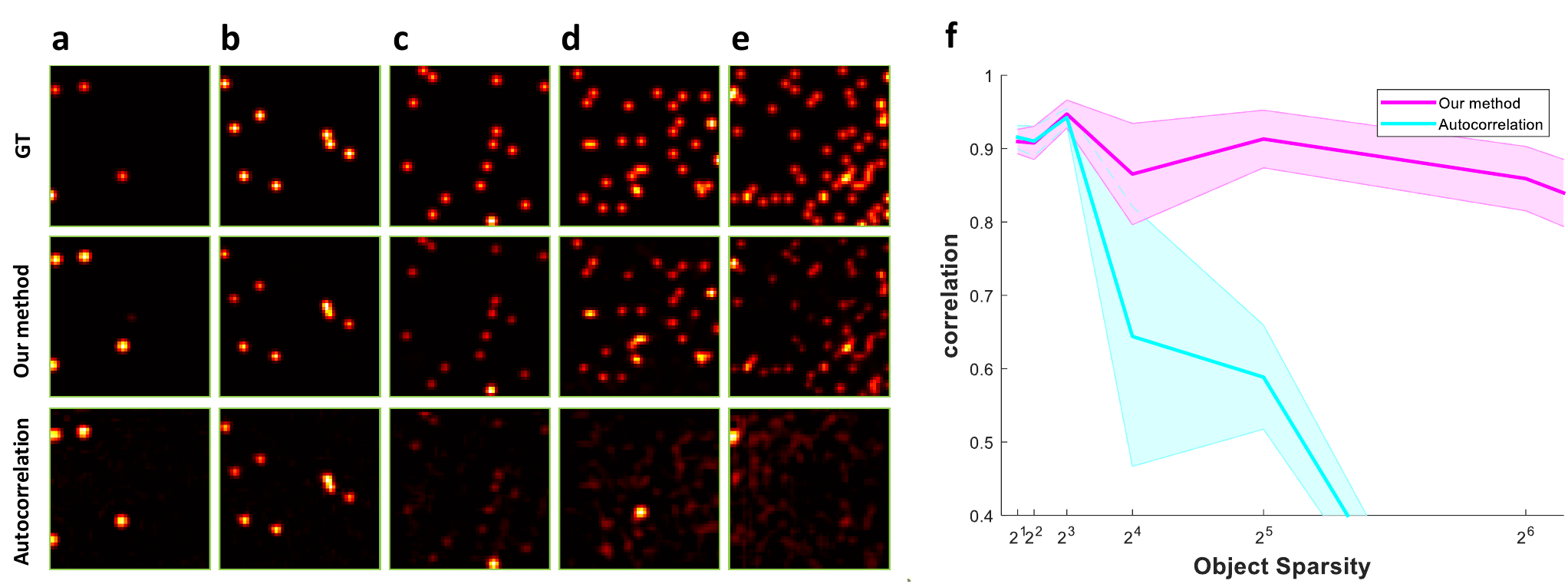}
\caption{\textbf{Numerical comparison between our method and autocorrelation-based reconstruction across varying object sparsity levels.}
(a-e) \textbf{Reconstruction results for objects with different sparsity levels.} Each column shows the ground truth, the reconstruction produced by our method, and the reconstruction obtained using the autocorrelation technique. (f) \textbf{Reconstruction fidelity versus object sparsity.} The correlation value indicates the correlation between the reconstruction result and the ground truth image, as a function of object sparsity. Each data point represents the statistical average from five independent numerical experiments using different randomly generated objects under identical conditions.}
\label{fig.6}
\end{figure*}

In terms of the imaging model and applicable scenarios, our imaging technique is similar to the autocorrelation work~\cite{bertolotti_non-invasive_2012,katz_non-invasive_2014,hofer_wide_2018,zhu_two-photon_2025}. However, unlike these methods, our approach achieves high-resolution reconstruction of complex objects by leveraging correlations within subimages, eliminating the need for phase retrieval. A key advantage of our technique is its superior performance on complex objects compared to autocorrelation methods.
To compare the performance of our method and autocorrelation technique on the different levels of the object, we present a numerical investigation of the reconstruction quality for various random objects with different sparsity in Fig.~\ref{fig.6}. The details of the simulation: The random objects are simulated by a random set of bright incoherent point fluorophores randomly dispersed in a $50 \times 50$ size grid with different sparsity levels. A randomly generated PSF of size $1280 \times 1280$ pixels is used, with a variance of 1.0 and its average intensity normalized to unity. The camera image size is $1231 \times 1231$ pixels. 150 camera images, which correspond to different random illuminations, are used in all the reconstruction procedures. For fairness, the autocorrelation-based reconstruction shown is the best result selected from 10 runs of the phase retrieval algorithm with different random phase initializations. Representative simulation results are displayed in Fig.~\ref{fig.6}a–e. The quantification curve as a function of object sparsity is shown in Fig.~\ref{fig.6}f.
The simulation results indicate that our method is more robust than the autocorrelation technique when reconstructing complex objects. 
It is worth noting that, under these simulation parameters, our method does not guarantee accurate reconstruction when the object sparsity exceeds $2^6$.

\section*{Acknowledgement}
This work was supported by FET OPEN DYNAMIC Chan Zuckerberg Initiative Deep Tissue Imaging; Agence Nationale de la Recherche (ANR-21-CE42-0013); H2020 Future and Emerging Technologies (863203); European Union’s Marie Skłodowska-Curie fellowship, grant agreement No. 888707 (DEEP3P). S.G. is supported by Institut Universitaire de France.

\bibliography{apssamp}% Produces the bibliography via BibTeX.

@article{bertolotti_imaging_2022,
	title = {Imaging in complex media},
	volume = {18},
	rights = {2022 Crown},
	issn = {1745-2481},
	url = {https://www.nature.com/articles/s41567-022-01723-8},
	doi = {10.1038/s41567-022-01723-8},
	pages = {1008--1017},
	number = {9},
	journal = {Nature Physics},
	author = {Bertolotti, Jacopo and Katz, Ori},
	urlyear = {2025-01-16},
	year = {2022},
	langid = {english},
}

@article{wu_single-shot_2016,
	title = {Single-shot diffraction-limited imaging through scattering layers via bispectrum analysis},
	volume = {41},
	issn = {0146-9592, 1539-4794},
	url = {https://www.osapublishing.org/abstract.cfm?URI=ol-41-21-5003},
	doi = {10.1364/OL.41.005003},
	number = {21},
	urldate = {2021-05-19},
	journal = {Optics Letters},
	author = {Wu, Tengfei and Katz, Ori and Shao, Xiaopeng and Gigan, Sylvain},
	year = {2016},
	pages = {5003},
}

@article{yoon_deep_2020,
	title = {Deep optical imaging within complex scattering media},
	volume = {2},
	rights = {2020 Springer Nature Limited},
	issn = {2522-5820},
	url = {https://www.nature.com/articles/s42254-019-0143-2},
	doi = {10.1038/s42254-019-0143-2},
	pages = {141--158},
	number = {3},
	journal = {Nature Reviews Physics},
	author = {Yoon, Seokchan and Kim, Moonseok and Jang, Mooseok and Choi, Youngwoon and Choi, Wonjun and Kang, Sungsam and Choi, Wonshik},
	year = {2020},
}

@article{helmchen_deep_2005,
	title = {Deep tissue two-photon microscopy},
	volume = {2},
	copyright = {2005 Nature Publishing Group},
	issn = {1548-7105},
	url = {https://www.nature.com/articles/nmeth818},
	number = {12},
	urldate = {2022-01-26},
	journal = {Nature Methods},
	author = {Helmchen, Fritjof and Denk, Winfried},
	year = {2005},
}

@article{wilson_resolution_2011,
	title = {Resolution and optical sectioning in the confocal microscope},
	volume = {244},
	copyright = {© 2011 The Author Journal of Microscopy © 2011 Royal Microscopical Society},
	issn = {1365-2818},
	url = {https://onlinelibrary.wiley.com/doi/abs/10.1111/j.1365-2818.2011.03549.x},
	number = {2},
	urldate = {2025-08-14},
	journal = {Journal of Microscopy},
	author = {Wilson, T.},
	year = {2011},
}

@article{hampson_adaptive_2021,
	title = {Adaptive optics for high-resolution imaging},
	volume = {1},
	copyright = {2021 Springer Nature Limited},
	issn = {2662-8449},
	url = {https://www.nature.com/articles/s43586-021-00066-7},
	number = {1},
	urldate = {2023-05-31},
	journal = {Nature Reviews Methods Primers},
	author = {Hampson, Karen M. and Turcotte, Raphaël and Miller, Donald T. and Kurokawa, Kazuhiro and Males, Jared R. and Ji, Na and Booth, Martin J.},
	year = {2021},
}

@article{feng_neuws_2023,
	title = {{NeuWS}: Neural wavefront shaping for guidestar-free imaging through static and dynamic scattering media},
	volume = {9},
	url = {https://www.science.org/doi/10.1126/sciadv.adg4671},
	doi = {10.1126/sciadv.adg4671},
	shorttitle = {{NeuWS}},
	pages = {eadg4671},
	number = {26},
	journal = {Science Advances},
	author = {Feng, Brandon Y. and Guo, Haiyun and Xie, Mingyang and Boominathan, Vivek and Sharma, Manoj K. and Veeraraghavan, Ashok and Metzler, Christopher A.},
	urldate = {2025-06-12},
	year = {2023},
}

@article{gigan_roadmap_2022,
	title = {Roadmap on wavefront shaping and deep imaging in complex media},
	volume = {4},
	issn = {2515-7647},
	url = {https://dx.doi.org/10.1088/2515-7647/ac76f9},
	doi = {10.1088/2515-7647/ac76f9},
	pages = {042501},
	number = {4},
	journal = {Journal of Physics: Photonics},
	author = {Gigan, Sylvain and Katz, Ori and de Aguiar, Hilton B and Andresen, Esben Ravn and Aubry, Alexandre and Bertolotti, Jacopo and Bossy, Emmanuel and Bouchet, Dorian and Brake, Joshua and Brasselet, Sophie and Bromberg, Yaron and Cao, Hui and Chaigne, Thomas and Cheng, Zhongtao and Choi, Wonshik and Čižmár, Tomáš and Cui, Meng and Curtis, Vincent R and Defienne, Hugo and Hofer, Matthias and Horisaki, Ryoichi and Horstmeyer, Roarke and Ji, Na and {LaViolette}, Aaron K and Mertz, Jerome and Moser, Christophe and Mosk, Allard P and Pégard, Nicolas C and Piestun, Rafael and Popoff, Sebastien and Phillips, David B and Psaltis, Demetri and Rahmani, Babak and Rigneault, Hervé and Rotter, Stefan and Tian, Lei and Vellekoop, Ivo M and Waller, Laura and Wang, Lihong and Weber, Timothy and Xiao, Sheng and Xu, Chris and Yamilov, Alexey and Yang, Changhuei and Yılmaz, Hasan},
	urlyear = {2025-01-16},
	year = {2022},
	langid = {english},
}

@article{sunray_beyond_2024,
	title = {Beyond memory-effect matrix-based imaging in scattering media by acousto-optic gating},
	volume = {9},
	issn = {2378-0967},
	url = {https://doi.org/10.1063/5.0219316},
	number = {9},
	urldate = {2025-08-13},
	journal = {APL Photonics},
	author = {Sunray, Elad and Weinberg, Gil and Rosenfeld, Moriya and Katz, Ori},
	year = {2024},
	pages = {096112},

}

@article{popoff_measuring_2010,
	title = {Measuring the Transmission Matrix in Optics: An Approach to the Study and Control of Light Propagation in Disordered Media},
	volume = {104},
	issn = {0031-9007, 1079-7114},
	url = {https://link.aps.org/doi/10.1103/PhysRevLett.104.100601},
	doi = {10.1103/PhysRevLett.104.100601},
	shorttitle = {Measuring the Transmission Matrix in Optics},
	pages = {100601},
	number = {10},
	journal = {Physical Review Letters},
	author = {Popoff, S. M. and Lerosey, G. and Carminati, R. and Fink, M. and Boccara, A. C. and Gigan, S.},
	urlyear = {2021-05-19},
	year = {2010},
	langid = {english},
}

@article{hsieh_imaging_2010,
	title = {Imaging through turbid layers by scanning the phase conjugated second harmonic radiation from a nanoparticle},
	volume = {18},
	rights = {\&\#169; 2010 {OSA}},
	issn = {1094-4087},
	url = {https://www.osapublishing.org/oe/abstract.cfm?uri=oe-18-20-20723},
	doi = {10.1364/OE.18.020723},
	pages = {20723--20731},
	number = {20},
	journal = {Optics Express},
	author = {Hsieh, Chia-Lung and Pu, Ye and Grange, Rachel and Laporte, Grégoire and Psaltis, Demetri},
	urlyear = {2021-09-07},
	year = {2010},
}

@article{aizik_non-invasive_2024,
	title = {Non-invasive and noise-robust light focusing using confocal wavefront shaping},
	volume = {15},
	copyright = {2024 The Author(s)},
	issn = {2041-1723},
	url = {https://www.nature.com/articles/s41467-024-49697-w},
	number = {1},
	urldate = {2025-08-11},
	journal = {Nature Communications},
	author = {Aizik, Dror and Levin, Anat},
	year = {2024},
}

@article{katz_looking_2012,
	title = {Looking around corners and through thin turbid layers in real time with scattered incoherent light},
	volume = {6},
	rights = {2012 Nature Publishing Group},
	issn = {1749-4893},
	url = {https://www.nature.com/articles/nphoton.2012.150},
	doi = {10.1038/nphoton.2012.150},
	pages = {549--553},
	number = {8},
	journal = {Nature Photonics},
	author = {Katz, Ori and Small, Eran and Silberberg, Yaron},
	urlyear = {2021-12-31},
	year = {2012},
	langid = {english},
	}

@article{aizik_fluorescent_2022,
	title = {Fluorescent wavefront shaping using incoherent iterative phase conjugation},
	volume = {9},
	copyright = {© 2022 Optica Publishing Group},
	issn = {2334-2536},
	url = {https://opg.optica.org/optica/abstract.cfm?uri=optica-9-7-746},
	number = {7},
	urldate = {2025-01-16},
	journal = {Optica},
	author = {Aizik, Dror and Gkioulekas, Ioannis and Levin, Anat},
	year = {2022},
	pages = {746--754},
}

@article{liu_directly_2023,
	title = {Directly and instantly seeing through random diffusers by self-imaging in scattering speckles},
	volume = {4},
	issn = {2662-1991},
	url = {https://doi.org/10.1186/s43074-022-00080-2},
	doi = {10.1186/s43074-022-00080-2},
	number = {1},
	urldate = {2026-02-26},
	journal = {PhotoniX},
	author = {Liu, Jietao and Yang, Wenhong and Song, Guofeng and Gan, Qiaoqiang},
	month = {jan},
	year = {2023},
	pages = {1},
}

@article{weinberg_noninvasive_2024,
	title = {Noninvasive megapixel fluorescence microscopy through scattering layers by a virtual incoherent reflection matrix},
	volume = {10},
	url = {https://www.science.org/doi/10.1126/sciadv.adl5218},
	doi = {10.1126/sciadv.adl5218},
	number = {47},
	urlyear = {2025-01-16},
	journal = {Science Advances},
	author = {Weinberg, Gil and Sunray, Elad and Katz, Ori},
	month = nov,
	year = {2024},
	pages = {eadl5218},
}

@article{yeminy_guidestar-free_2021,
	title = {Guidestar-free image-guided wavefront shaping},
	volume = {7},
	copyright = {Copyright © 2021 The Authors, some rights reserved; exclusive licensee American Association for the Advancement of Science. No claim to original U.S. Government Works. Distributed under a Creative Commons Attribution NonCommercial License 4.0 (CC BY-NC).. https://creativecommons.org/licenses/by-nc/4.0/This is an open-access article distributed under the terms of the Creative Commons Attribution-NonCommercial license, which permits use, distribution, and reproduction in any medium, so long as the resultant use is not for commercial advantage and provided the original work is properly cited.},
	issn = {2375-2548},
	url = {https://advances.sciencemag.org/content/7/21/eabf5364},
	doi = {10.1126/sciadv.abf5364},
	number = {21},
	urldate = {2021-05-26},
	journal = {Science Advances},
	author = {Yeminy, Tomer and Katz, Ori},
	month = {may},
	year = {2021},
	note = {Publisher: American Association for the Advancement of Science
Section: Research Article},
	pages = {eabf5364},

}

@article{boniface_non-invasive_2020,
	title = {Non-invasive focusing and imaging in scattering media with a fluorescence-based transmission matrix},
	volume = {11},
	issn = {2041-1723},
	url = {http://arxiv.org/abs/2003.04255},
	doi = {10.1038/s41467-020-19696-8},
	number = {1},
	urlyear = {2021-05-19},
	journal = {Nature Communications},
	author = {Boniface, Antoine and Dong, Jonathan and Gigan, Sylvain},
	month = dec,
	year = {2020},
	pages = {6154},
}

@article{may_fast_2021,
	title = {Fast holographic scattering compensation for deep tissue biological imaging},
	volume = {12},
	copyright = {2021 The Author(s)},
	issn = {2041-1723},
	url = {https://www.nature.com/articles/s41467-021-24666-9},
	number = {1},
	urldate = {2025-06-15},
	journal = {Nature Communications},
	author = {May, Molly A. and Barré, Nicolas and Kummer, Kai K. and Kress, Michaela and Ritsch-Marte, Monika and Jesacher, Alexander},
	year = {2021},
	pages = {4340},

}

@article{darco_physics-based_2022,
	title = {Physics-based neural network for non-invasive control of coherent light in scattering media},
	volume = {30},
	copyright = {© 2022 Optica Publishing Group},
	issn = {1094-4087},
	url = {https://opg.optica.org/oe/abstract.cfm?uri=oe-30-17-30845},
	doi = {10.1364/OE.465702},
	number = {17},
	urlyear = {2025-01-16},
	journal = {Optics Express},
	author = {d’Arco, Alexandra and Xia, Fei and Boniface, Antoine and Dong, Jonathan and Gigan, Sylvain},
	month = {aug},
	year = {2022},
	pages = {30845--30856},
}

@article{daniel_light_2019,
	title = {Light focusing through scattering media via linear fluorescence variance maximization, and its application for fluorescence imaging},
	volume = {27},
	issn = {1094-4087},
	url = {https://www.osapublishing.org/abstract.cfm?URI=oe-27-15-21778},
	number = {15},
	urldate = {2021-05-19},
	journal = {Optics Express},
	author = {Daniel, Anat and Oron, Dan and Silberberg, Yaron},
	year = {2019},
}

@article{berlage_deep_2021,
	title = {Deep tissue scattering compensation with three-photon {F}-{SHARP}},
	volume = {8},
	copyright = {© 2021 Optical Society of America},
	issn = {2334-2536},
	url = {https://opg.optica.org/optica/abstract.cfm?uri=optica-8-12-1613},
	number = {12},
	urldate = {2025-08-10},
	journal = {Optica},
	author = {Berlage, Caroline and Tantirigama, Malinda L. S. and Babot, Mathias and Battista, Diego Di and Whitmire, Clarissa and Papadopoulos, Ioannis N. and Poulet, James F. A. and Larkum, Matthew and Judkewitz, Benjamin},
	year = {2021},
}

@article{zhao_single-pixel_2024,
	title = {Single-pixel transmission matrix recovery via two-photon fluorescence},
	volume = {10},
	url = {https://www.science.org/doi/10.1126/sciadv.adi3442},
	doi = {10.1126/sciadv.adi3442},
	number = {3},
	urlyear = {2025-01-16},
	journal = {Science Advances},
	author = {Zhao, Shupeng and Rauer, Bernhard and Valzania, Lorenzo and Dong, Jonathan and Liu, Ruifeng and Li, Fuli and Gigan, Sylvain and de Aguiar, Hilton B.},
	month = jan,
	year = {2024},
	pages = {eadi3442},
}

@article{papadopoulos_scattering_2017,
	title = {Scattering compensation by focus scanning holographic aberration probing ({F}-{SHARP})},
	volume = {11},
	copyright = {2016 Nature Publishing Group},
	issn = {1749-4893},
	url = {https://www.nature.com/articles/nphoton.2016.252},
	doi = {10.1038/nphoton.2016.252},
	number = {2},
	urlyear = {2021-10-25},
	journal = {Nature Photonics},
	author = {Papadopoulos, Ioannis N. and Jouhanneau, Jean-Sébastien and Poulet, James F. A. and Judkewitz, Benjamin},
	year = {2017},
	pages = {116--123},
}

@article{boniface_non-invasive_2019,
	title = {Non-invasive light focusing in scattering media using speckle variance optimization},
	volume = {6},
	issn = {2334-2536},
	url = {http://arxiv.org/abs/1906.01574},
	doi = {10.1364/OPTICA.6.001381},
	number = {11},
	urldate = {2021-08-04},
	journal = {Optica},
	author = {Boniface, Antoine and Blochet, Baptiste and Dong, Jonathan and Gigan, Sylvain},
	month = {nov},
	year = {2019},
	pages = {1381},
}

@article{zhu_large_2022,
	title = {Large field-of-view non-invasive imaging through scattering layers using fluctuating random illumination},
	volume = {13},
	rights = {2022 The Author(s)},
	issn = {2041-1723},
	url = {https://www.nature.com/articles/s41467-022-29166-y},
	doi = {10.1038/s41467-022-29166-y},
	pages = {1447},
	number = {1},
	journal = {Nature Communications},
	author = {Zhu, Lei and Soldevila, Fernando and Moretti, Claudio and d’Arco, Alexandra and Boniface, Antoine and Shao, Xiaopeng and de Aguiar, Hilton B. and Gigan, Sylvain},
	urlyear = {2022-03-18},
	year = {2022},
	langid = {english},
}

@article{baek_phase_2023,
	title = {Phase conjugation with spatially incoherent light in complex media},
	volume = {17},
	rights = {2023 The Author(s), under exclusive licence to Springer Nature Limited},
	issn = {1749-4893},
	url = {https://www.nature.com/articles/s41566-023-01254-5},
	doi = {10.1038/s41566-023-01254-5},
	pages = {1114--1119},
	number = {12},
	journal = {Nature Photonics},
	author = {Baek, {YoonSeok} and de Aguiar, Hilton B. and Gigan, Sylvain},
	urlyear = {2024-01-22},
	year = {2023},
	langid = {english},
}

@misc{lewis1995fncc,
  author       = {Lewis, J. P.},
  title        = {Fast Normalized Cross-Correlation},
  year         = {1995},
  howpublished = {\url{https://scribblethink.org/Work/nvisionInterface/nip.pdf}},
  note         = {Industrial Light \& Magic}
}

@article{zhu_two-photon_2025,
	title = {Two-photon microscopy through scattering media harnessing speckle autocorrelation},
	volume = {33},
	issn = {1094-4087},
	url = {https://opg.optica.org/oe/abstract.cfm?uri=oe-33-25-52860},
	doi = {10.1364/OE.567297},
	number = {25},
	urldate = {2025-12-10},
	journal = {Optics Express},
	author = {Zhu, Lei and Rauer, Bernhard and Aguiar, Hilton B. de and Gigan, Sylvain},
	month = {dec},
	year = {2025},
	pages = {52860--52867},

}

@article{freund_memory_1988,
	title = {Memory {Effects} in {Propagation} of {Optical} {Waves} through {Disordered} {Media}},
	volume = {61},
	issn = {0031-9007},
	url = {https://link.aps.org/doi/10.1103/PhysRevLett.61.2328},
	number = {20},
	urldate = {2021-05-19},
	journal = {Physical Review Letters},
	author = {Freund, Isaac and Rosenbluh, Michael and Feng, Shechao},
	year = {1988},
}

@article{katz_non-invasive_2014,
	title = {Non-invasive single-shot imaging through scattering layers and around corners via speckle correlations},
	volume = {8},
	rights = {2014 Nature Publishing Group},
	issn = {1749-4893},
	url = {https://www.nature.com/articles/nphoton.2014.189},
	doi = {10.1038/nphoton.2014.189},
	pages = {784--790},
	number = {10},
	journal  = {Nature Photonics},
	author = {Katz, Ori and Heidmann, Pierre and Fink, Mathias and Gigan, Sylvain},
	urlyear = {2021-05-19},
	year = {2014},
	langid = {english},
}

@article{bertolotti_non-invasive_2012,
	title = {Non-invasive imaging through opaque scattering layers},
	volume = {491},
	issn = {0028-0836, 1476-4687},
	url = {http://www.nature.com/articles/nature11578},
	doi = {10.1038/nature11578},
	pages = {232--234},
	number = {7423},
	journal = {Nature},
	author = {Bertolotti, Jacopo and van Putten, Elbert G. and Blum, Christian and Lagendijk, Ad and Vos, Willem L. and Mosk, Allard P.},
	urlyear = {2021-05-19},
	year = {2012},
	langid = {english},
}

@article{tian_single-shot_2022,
	title = {Single-shot imaging through scattering medium with a deterministic phase-retrieval algorithm},
	volume = {506},
	issn = {0030-4018},
	url = {https://www.sciencedirect.com/science/article/pii/S0030401821008117},
	doi = {10.1016/j.optcom.2021.127562},
	pages = {127562},
	journal = {Optics Communications},
	author = {Tian, Bingxin and Zhu, Lei and Liu, Bingcai and Han, Jun},
	urlyear = {2021-10-31},
	year = {2022},
	langid = {english},
}

@article{hofer_wide_2018,
	title = {Wide field fluorescence epi-microscopy behind a scattering medium enabled by speckle correlations},
	volume = {26},
	issn = {1094-4087},
	url = {https://www.osapublishing.org/oe/abstract.cfm?uri=oe-26-8-9866},
	doi = {10.1364/OE.26.009866},
	pages = {9866--9881},
	number = {8},
	journal = {Optics Express},
	author = {Hofer, Matthias and Soeller, Christian and Brasselet, Sophie and Bertolotti, Jacopo},
	urlyear = {2021-05-19},
	year = {2018},
}

@article{freund_looking_1990,
	title = {Looking through walls and around corners},
	volume = {168},
	issn = {0378-4371},
	url = {https://www.sciencedirect.com/science/article/pii/037843719090357X},
	number = {1},
	urldate = {2025-09-12},
	journal = {Physica A: Statistical Mechanics and its Applications},
	author = {Freund, Isaac},
	year = {1990},
	pages = {49--65},
}

@article{biggs_acceleration_1997,
	title = {Acceleration of iterative image restoration algorithms},
	volume = {36},
	issn = {0003-6935, 1539-4522},
	url = {https://www.osapublishing.org/abstract.cfm?URI=ao-36-8-1766},
	number = {8},
	urldate = {2021-02-15},
	journal = {Applied Optics},
	author = {Biggs, David S. C. and Andrews, Mark},
	month = {mar},
	year = {1997},
	pages = {1766},
}

@article{basak_super-resolution_2025,
	title = {Super-resolution optical fluctuation imaging},
	volume = {19},
	copyright = {2025 Springer Nature Limited},
	issn = {1749-4893},
	url = {https://www.nature.com/articles/s41566-024-01571-3},
	doi = {10.1038/s41566-024-01571-3},
	number = {3},
	urldate = {2025-08-20},
	journal = {Nature Photonics},
	author = {Basak, Samrat and Chizhik, Alexey and Gallea, José Ignacio and Gligonov, Ivan and Gregor, Ingo and Nevskyi, Oleksii and Radmacher, Niels and Tsukanov, Roman and Enderlein, Jörg},
	month = {mar},
	year = {2025},
	pages = {229--237},
}

@article{fienup_phase_1982,
	title = {Phase retrieval algorithms: a comparison},
	volume = {21},
	copyright = {© 1982 Optical Society of America},
	issn = {2155-3165},
	shorttitle = {Phase retrieval algorithms},
	url = {https://opg.optica.org/ao/abstract.cfm?uri=ao-21-15-2758},
	number = {15},
	urldate = {2024-10-04},
	journal = {Applied Optics},
	author = {Fienup, J. R.},
	year = {1982},
	pages = {2758--2769},

}

@article{osnabrugge_generalized_2017,
	title = {Generalized optical memory effect},
	volume = {4},
	rights = {\&\#169; 2017 Optical Society of America},
	issn = {2334-2536},
	url = {https://www.osapublishing.org/optica/abstract.cfm?uri=optica-4-8-886},
	doi = {10.1364/OPTICA.4.000886},
	pages = {886--892},
	number = {8},
	journal = {Optica},
	author = {Osnabrugge, Gerwin and Horstmeyer, Roarke and Papadopoulos, Ioannis N. and Judkewitz, Benjamin and Vellekoop, Ivo M.},
	urlyear = {2021-05-20},
	year = {2017},
}

@article{soldevila_functional_2023,
	title = {Functional imaging through scattering medium via fluorescence speckle demixing and localization},
	volume = {31},
	rights = {© 2023 Optica Publishing Group},
	issn = {1094-4087},
	url = {https://opg.optica.org/oe/abstract.cfm?uri=oe-31-13-21107},
	doi = {10.1364/OE.487768},
	pages = {21107--21117},
	number = {13},
	journal = {Optics Express},
	author = {Soldevila, F. and Moretti, C. and Nöbauer, T. and Sarafraz, H. and Vaziri, A. and Gigan, S.},
	urlyear = {2025-01-20},
	year = {2023},
}

@article{wang_non-invasive_2021,
	title = {Non-invasive super-resolution imaging through dynamic scattering media},
	volume = {12},
	copyright = {2021 The Author(s)},
	issn = {2041-1723},
	url = {https://www.nature.com/articles/s41467-021-23421-4},
	number = {1},
	urldate = {2021-06-14},
	journal = {Nature Communications},
	author = {Wang, Dong and Sahoo, Sujit K. and Zhu, Xiangwen and Adamo, Giorgio and Dang, Cuong},
	year = {2021},
}

@article{wu_replica-assisted_2025,
	title = {Replica-{Assisted} {Super}-{Resolution} {Fluorescence} {Imaging} in {Scattering} {Media}},
	volume = {12},
	url = {https://doi.org/10.1021/acsphotonics.4c02468},
	number = {3},
	urldate = {2025-08-12},
	journal = {ACS Photonics},
	author = {Wu, Tengfei and Baek, YoonSeok and Xia, Fei and Gigan, Sylvain and de Aguiar, Hilton B.},
	year = {2025},
}

@article{dertinger_fast_2009,
	title = {Fast, background-free, {3D} super-resolution optical fluctuation imaging ({SOFI})},
	volume = {106},
	copyright = {© 2009},
	issn = {0027-8424, 1091-6490},
	url = {https://www.pnas.org/content/106/52/22287},
	doi = {10.1073/pnas.0907866106},
	number = {52},
	journal = {Proceedings of the National Academy of Sciences},
	author = {Dertinger, T. and Colyer, R. and Iyer, G. and Weiss, S. and Enderlein, J.},
	month = {dec},
	year = {2009},
	pmid = {20018714},
	pages = {22287--22292},

}

\end{document}